\documentclass[sigconf,nonacm]{acmart}
\renewcommand\footnotetextcopyrightpermission[1]{} 
\usepackage{graphicx} 
\usepackage{subfigure}
\usepackage{stfloats} 
\usepackage{marvosym}

\begin{document}

\title{Think-on-Process: Dynamic Process Generation for Collaborative Development of Multi-Agent System}



\author{Leilei Lin}
\authornote{Equal Contribution.}
\email{leilei_lin@126.com}
\affiliation{%
  \institution{Capital Normal University}
  \city{Beijing}
  \country{China}
}

\author{Yingming Zhou}
\authornotemark[1]
\email{aizhouym@gmail.com}
\affiliation{%
  \institution{Capital Normal University}
  \city{Beijing}
  \country{China}
}

\author{Wenlong Chen}
\email{wenlongchen@sina.com}
\affiliation{%
  \institution{Capital Normal Univeristy}
  \city{Beijing}
  \country{China}
}

\author{Chen Qian}
\authornote{Corresponding Author.}
\email{qianc62@gmail.com}
\affiliation{%
  \institution{Tsinghua University}
  \city{Beijing}
  \country{China}
}

\begin{abstract}
Software development is a collaborative endeavor that requires individuals from different departments to work together in order to collectively develop a high-quality software system. In this context, people have begun to explore a method that leverages multi-agent systems based on LLMs to carry out software development. However, existing research tends to rigidly fix the software development process in a framework in code form, thus failing to dynamically adjust the software development process in real-time to meet the more flexible and variable software environment. In this paper, we propose a dynamic process generation framework, named \textbf{ToP (Think-on-Process)}. The core idea of ToP is to leverage experiential knowledge (i.e., process models) to guide LLMs in generating software development processes (i.e., instances).  These instances will guide multi-agent in software development and employ a compiler to provide feedback on the development outcomes. Subsequently, we utilize heuristic algorithms to filter the instances and apply process mining algorithms to derive process model. Finally, the process model will be converted into text, formatted as prompts, to enhance the ability of LLMs to generate other instances. Experiments demonstrate that our framework ToP significantly enhances the dynamic process generation capability of the GPT-3.5 and GPT-4 for five categories of software development tasks. 
\end{abstract}


\maketitle

\section{INTRODUCTION}

In the software development lifecycle, errors that occur during the development phases can result in substantial resource wastage and a decline in workforce efficiency. This can be particularly harmful to startups and small businesses. To maintain the integrity and tractability of the software development phases, numerous companies tackle these challenges by adopting structured software development processes \cite{coleman2008investigation,clarke2012situational}. These processes outline the necessary steps that a development team should take during each production phase, and they strictly manage various facets, including planning, quality assessment, and scheduling. By strictly adhering to a structured development process, project managers can effectively oversee the project’s progress, enabling the early identification and rectification of potential issues. This proactive approach helps to avert future project setbacks and avoidable financial losses \cite{paz2016systematic,shylesh2017study}.

\begin{figure}[t]
\centering
\centerline{\includegraphics[scale=0.34]{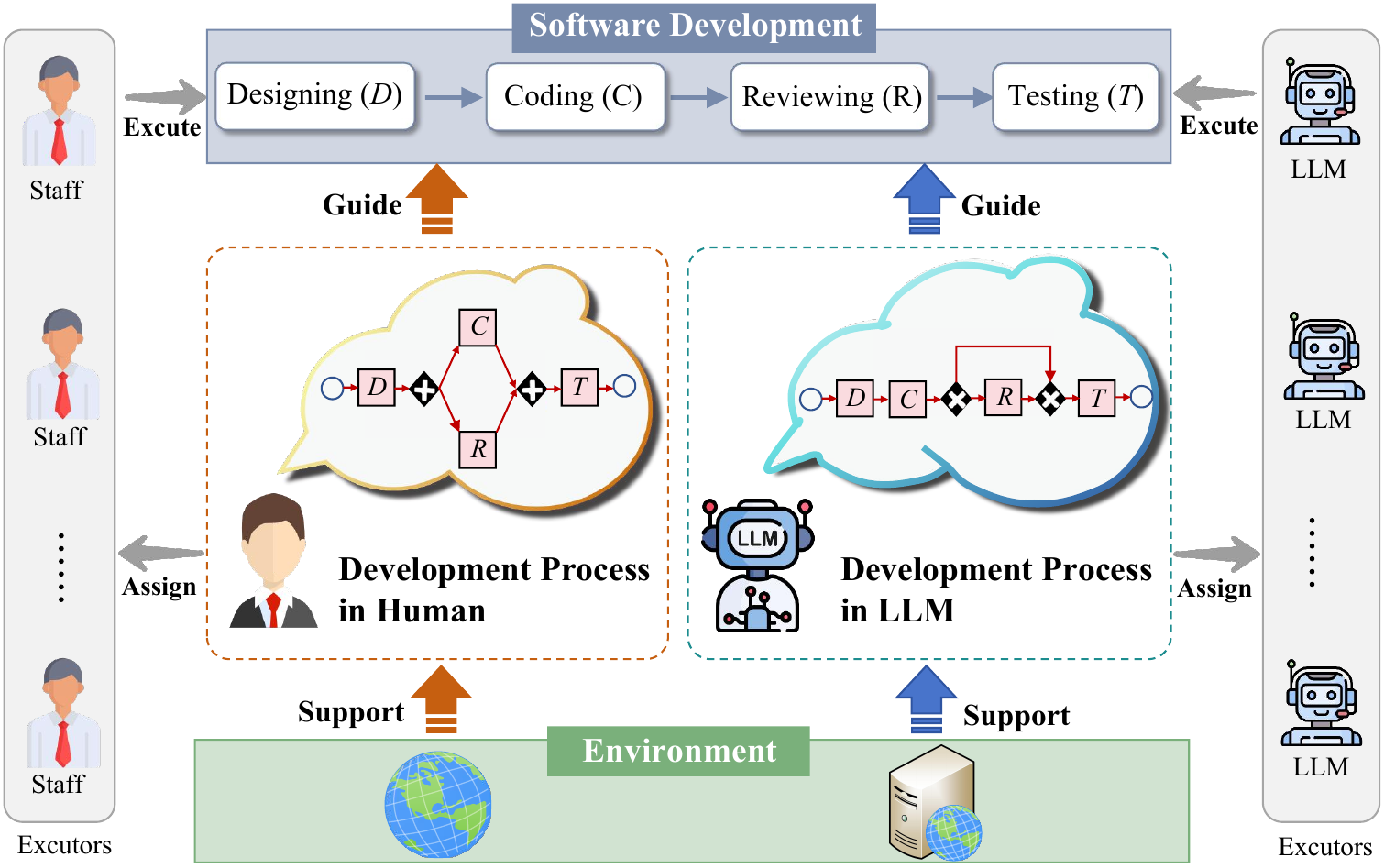}}
\caption{Traditional software development  and agent-based software development.}
\label{fig1}
\end{figure}

\begin{figure*}[h]
\centering
\centerline{\includegraphics[scale=0.48]{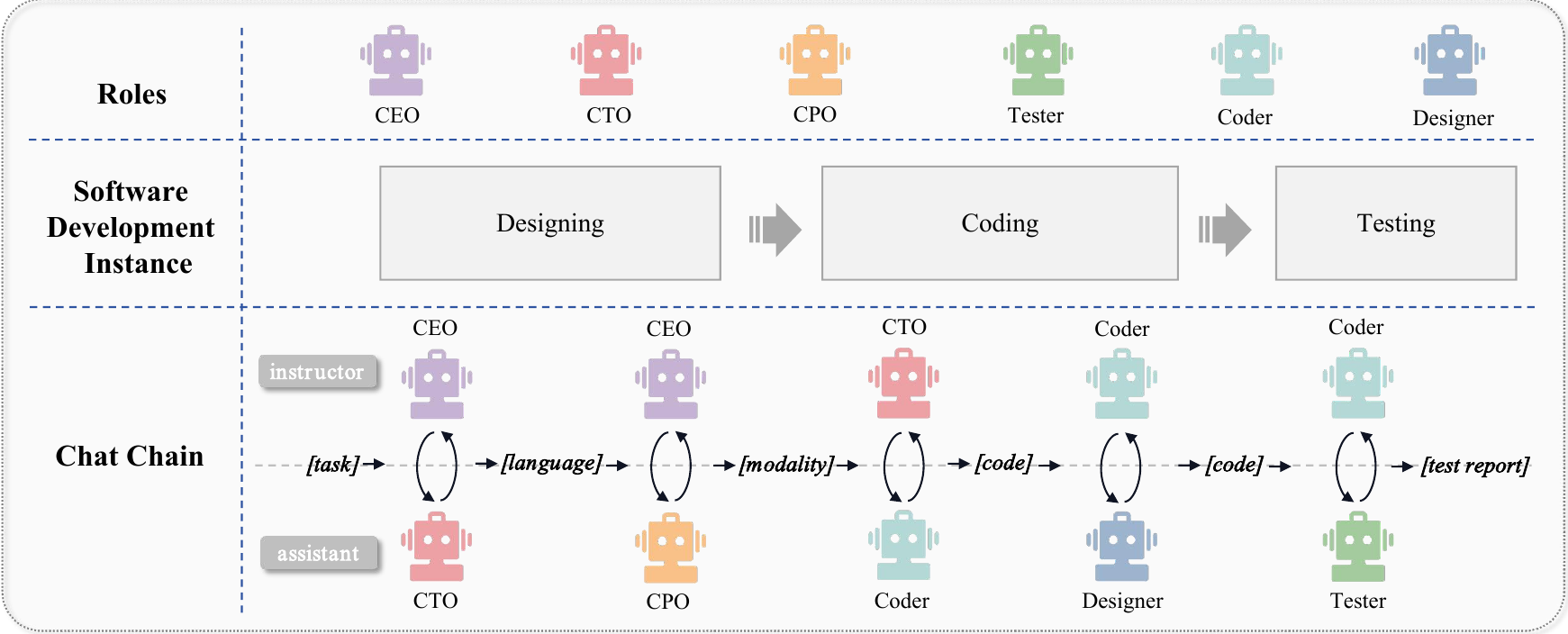}}
\caption{The Chat Chain in Multi-Agent Software Development.}
\label{fig15}
\end{figure*}

The central concept of a multi-agent system is to decompose complex problems into several distinct subtasks, which are then allocated to autonomous agents within the system for processing \cite{dorri2018multi}. By facilitating communication and collaboration among these agents, the system can effectively solve the original complex problems. Multi-agent systems continue to attract scholarly interest across various disciplines due to their adaptability in addressing complex issues and their potential for intelligence \cite{xie2017multi}. In recent years, Large Language Models (LLMs) have made substantial advancements in the field of Natural Language Processing (NLP) \cite{brown2020language,touvron2023llama,bubeck2023sparks,park2023generative}, leveraging massive corpora for training with the objective of ``next word prediction.'' LLMs have exhibited remarkable performance across a spectrum of downstream tasks. For instance, Ni et al. \cite{ni2023chatreport} have employed LLMs as automated analysis tools for continuous report analysis, Nijikam et al. \cite{nijkamp2022codegen} have used them to generate executable code through iterative dialogue, and Cheng et al. \cite{cheng2023prompt} have harnessed LLMs to create software engineering infrastructure that empowers individuals to innovate with AI. The question arises: How can we harness the full potential of LLMs in software development? In essence, it involves creating a network of agents that collaborate to execute tasks, which primes us to anticipate the performance of LLM-powered multi-agent system in software development with great interest. As depicted in Figure \ref{fig1}, two fundamentally different approaches to software development are contrasted. Conventional software development relies on human expertise, defining the development process in advance, and tasking different departmental staff with specific phase tasks under the leadership's guidance. In contrast, an LLM-powered multi-agent system leverages the collaboration of multiple autonomous agents to navigate the entire software development lifecycle without human intervention. This method achieves a level of automation in software development that far surpasses traditional models, significantly reducing labor requirements.

Researchers have begun to integrate multi-agent system into software development processes. Qian et al. \cite{qian2023communicative} introduced ChatDev, a virtual chat-based software technology company capable of automating software development according to the waterfall model. Qian et al. \cite{qian2023experiential} enhanced the capability of autonomous agents by enabling them to gather shortcut experiences from historical trajectories. These past experiences are then employed for mutual inference, which decreases the recurrence of repetitive errors in software development. Despite these advancements, the current framework for software development utilizing multi-agent system faces limitations, notably its static nature. This implies that the phases in the software development process are immutable, which hampers its ability to accommodate the flexible and evolving demands of modern development. A dynamic software development process, however, can adapt to various development scenarios and optimize resource utilization while satisfying diverse user requirements. Consequently, there is a need to investigate a dynamic process generation framework that can facilitate real-time collaboration within a multi-agent system during software development. Our contributions are as follows:

\begin{itemize}
    \item We have developed an architecture capable of dynamically generating various process instances to facilitate software development across diverse scenarios. This architecture assigns autonomous agents to collaborate on development tasks based on these instances.
    \item To address potential hallucination issues that LLMs may encounter during instance generation, we use temperature settings and external environment (e.g. compiler) to filter out erroneous instances (referring to those with incorrect development phases, unreasonable order of development phases, or both, as well as software compilation errors guided by instances), and use this mechanism to retain high-quality and executable instances.
    \item We utilize process mining algorithms to extract the software development process from successful instances. These process is then transformed into textual descriptions, which used to augment the dynamic instance generation capabilities of LLMs. The efficacy of our approach has been validated through rigorous experimental analysis.
\end{itemize}

\begin{figure*}[htbp]
\centering
\centerline{\includegraphics[scale=0.45]{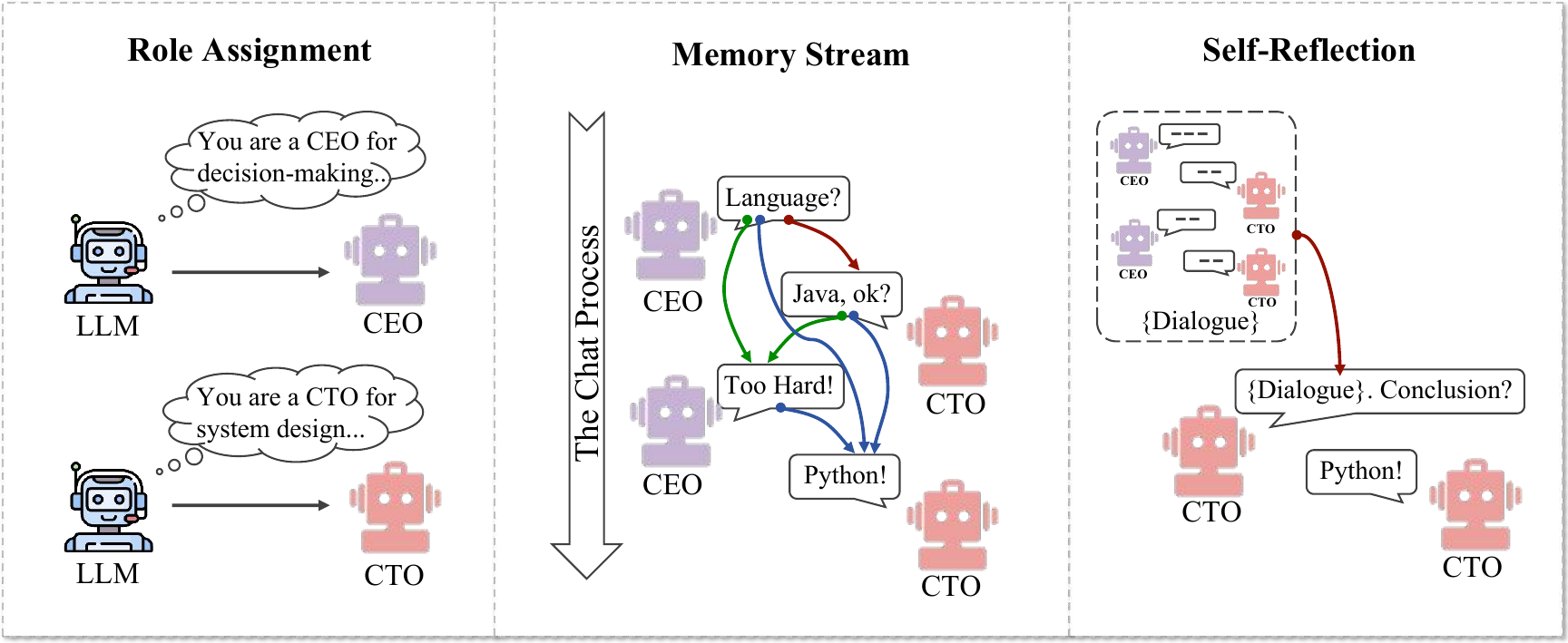}}
\caption{Three mechanisms used in every atomic chat.}
\label{fig16}
\end{figure*}

\section{BACKGROUND}
Research into guiding multi-agent software development with static processes is extensive and well-documented. In the following section, we will present ChatDev \cite{qian2023communicative}, a virtual chat-based software technology company. ChatDev comprises agents with distinct roles that collaborate to develop the necessary software. Additionally, we will delve into the Chat Chain and the three mechanisms employed within ChatDev to gain insight into how the company addresses the challenges of missing dependencies and index out-of-bounds errors that LLMs might face when generating comprehensive software systems.

\textbf{ Chat Chain}: Common software development life cycles encompass models such as the waterfall model, V-model, spiral model, and agile development \cite{balaji2012waterfall,alshamrani2015comparison,akinsola2020comparative}. These models represent distinct approaches to managing the software development process. ChatDev employs the waterfall model, which structures software development into four phases: Designing, Coding, Testing, and Documenting. The strategic allocation of roles for collaborative development within each phase is crucial to the overall success of the software. The Chat Chain mechanism employed by ChatDev breaks down each phase into multiple atomic chats. In each atomic chat, two agents with predefined roles take on the roles of instructor and assistant, respectively, to engage in focused communication and task execution, ultimately achieving the desired output. Figure \ref{fig15} illustrates a software development instance comprising three phases: Designing, Coding, and Testing. Each phase is executed through one or two atomic chats. By sequentially completing these atomic chats, the collaborative development of the software is brought to fruition. The Chat Chain's structure provides users with a clear view of every detailed task involved in software development, facilitating subsequent inspections and enhancements. Moreover, the division of software development into smaller subtasks through atomic chats significantly mitigates the potential for code hallucinations, enhancing the overall reliability of the development process.

Having grasped how ChatDev disaggregates software development phases into various atomic chats via the Chat Chain mechanism, we now turn our attention to the three mechanisms employed within each chat, as depicted in Figure \ref{fig16}:

\textit{Role Assignment}: Prompts are recognized as a critical tool to steer LLMs in the desired direction, with their efficacy substantiated by numerous studies \cite{reynolds2021prompt,zhou2022large,white2023prompt}. In role-playing scenarios, prompts containing details such as assigned tasks, roles, output specifications, and communication protocols are provided to different agents. This ensures that each agent comprehends and fulfills its designated responsibilities.

\textit{Memory Stream}: The memory stream mechanism \cite{park2023generative} involves the thorough recording of past communication between agents. This historical dialogue data aids agents in decision-making during new dialogues, preventing the erosion of past experiences \cite{zhang2024survey}. Additionally, by establishing a communication protocol, agents are assured of delivering terminal messages in a consistent format upon reaching a consensus.

\textit{Self-Reflection}: Occasionally, the assistant's decisions may not align with the communication protocol even when a consensus is reached. This can lead to outputs in subsequent development phases that fall short of expectations. Self-reflection addresses this issue by prompting the assistant to revisit past dialogues (effectively initiating a new dialogue round based on historical context). This process facilitates the derivation of conclusions that adhere to the established communication protocol. These mechanisms collectively enhance the effectiveness and reliability of the collaborative development process managed by ChatDev.

\begin{figure*}[h]
\centering
\centerline{\includegraphics[scale=0.65]{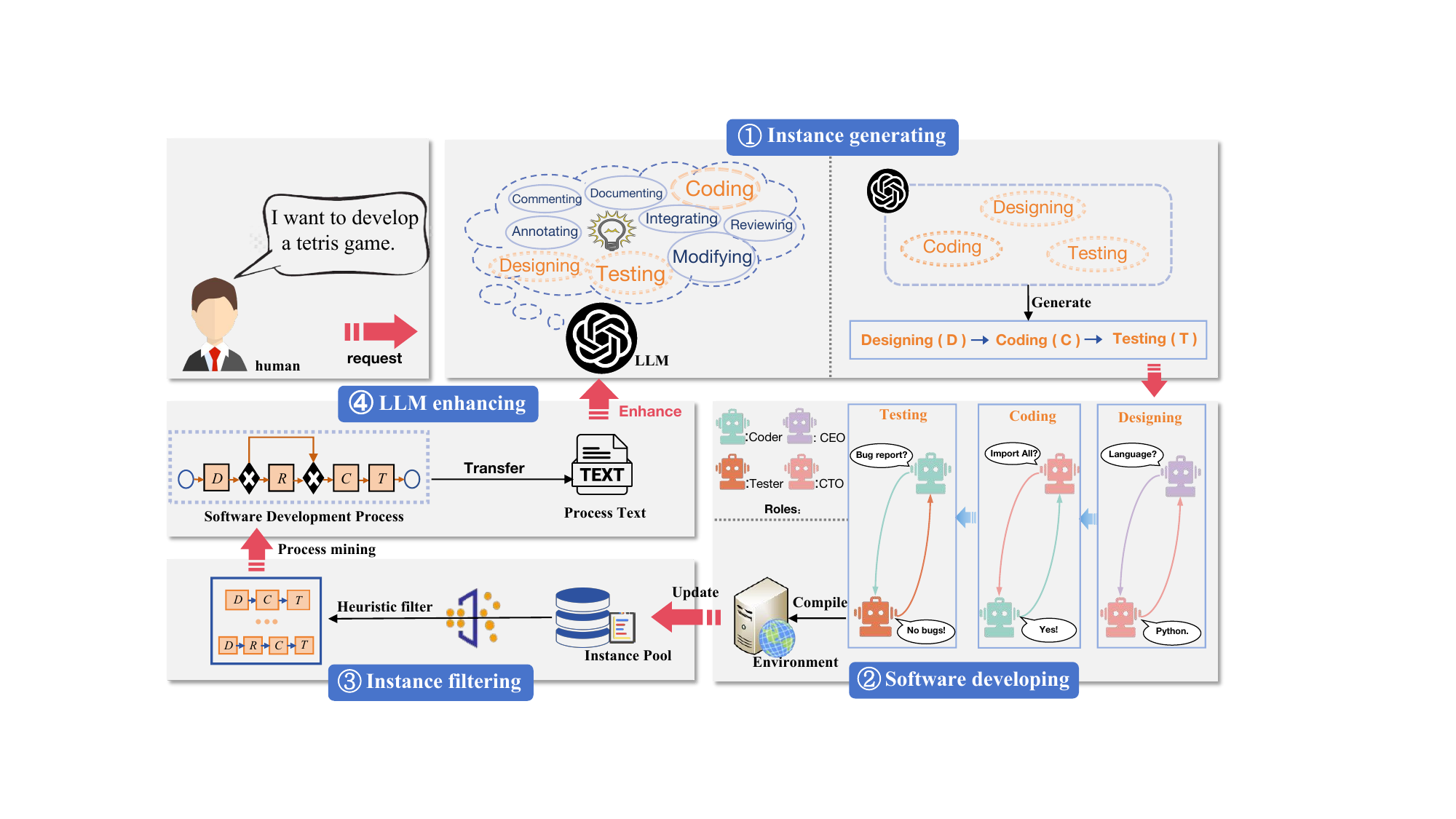}}
\caption{ An overview of our method named ToP.}
\label{fig2}
\end{figure*}

\section{METHODOLOGY}
By dynamically generating the development process, our framework can effectively meet the adaptable and evolving needs of software development, while also providing guidance for ongoing multi-agent collaboration. As depicted in Figure \ref{fig2}, we will outline our framework through four stages: Instance generating, Software developing, Instance filtering, and LLMs enhancing.

\subsection{Instance generating}
Dynamic process generation involves the creation of various instances based on LLMs' existing knowledge of software development, tailored to the specific and varied needs of users. These instances serve as blueprints to direct the activities of multiple agents involved in software development. It is essential to recognize that software development processes can differ significantly depending on the project requirements. Consequently, adhering to a single, static development process is not always feasible or efficient. Our proposal for dynamic process generation aims to accommodate this variability, enabling a wider array of diverse instances to emerge and guide the development process accordingly. The concept of instances and diverse instances is explained as follows:
\begin{itemize}
    \item Instance: An instance represents a software development process that includes different phases of software development. These phases are arranged in chronological order. As shown in Figure \ref{fig2}, ``\text{Designing (D)} $\to$ \text{Coding (C)} $\to$ \text{Testing (T)}'' is an instance.
    \item Diverse instances: Diverse instances refer to multiple instances that are pairwise different from each other, including different phases, orders, or both. For example, ``D $\to$ C $\to$ T'', ``D $\to$ R $\to$ C $\to$ T'', and ``D $\to$ C $\to$ R $\to$T'' are diverse instances.
\end{itemize}

\begin{figure}[b]
\centering
\centerline{\includegraphics[scale=0.44]{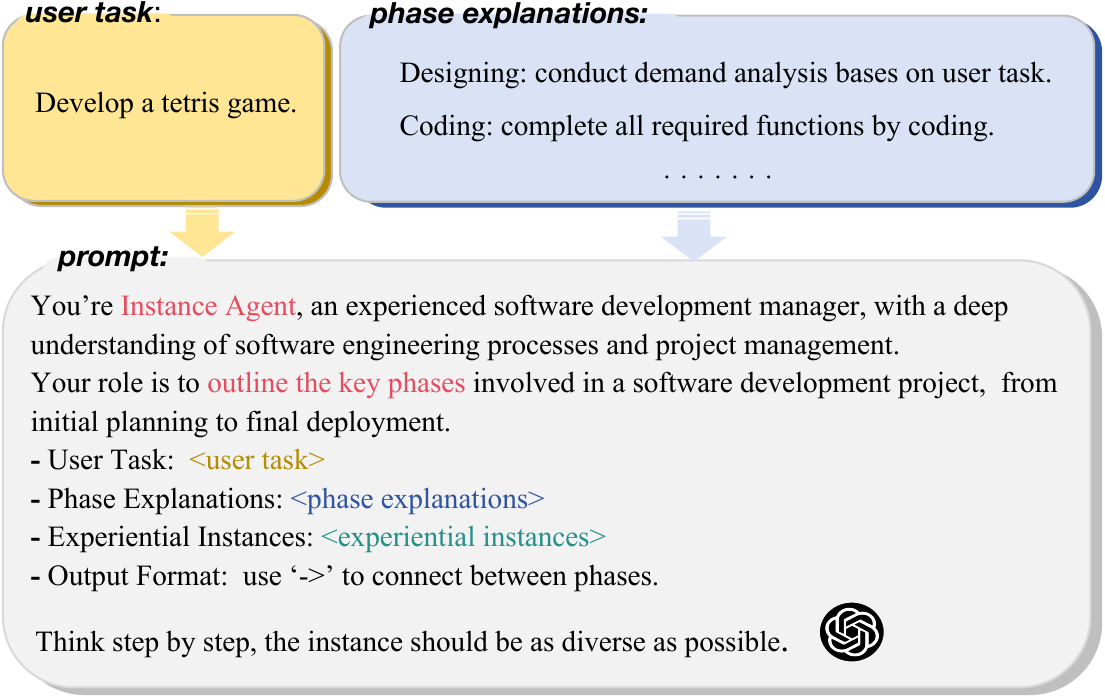}}
\caption{The prompts for LLMs to generate instances.}
\label{fig3}
\end{figure}

As shown in \ref{fig3}, when LLMs receives a development task similar to ``Develop a tetris game,'' we use the prompt engineering to prompt it to generate diverse instances. The core components of prompt includes four parts: 
1) \textit {User Task}: Provided by the user, which is then passed as a prompt to LLMs; 
2) \textit{Phase Explanations}: The phase explanations provide a detailed description of each phase and the development tasks that need to be completed, ensuring that LLMs can generate the correct phases in subsequent instance generation; 
3) \textit{Experiential Instances}: To better guide LLMs in accurately generating instances, provide some existing instances each time; 
4) \textit {Output Format}: Using the special symbol `$\to$' to connect different phases into a complete instance. LLMs assigns probabilities to all possible tokens based on its understanding of the entire context. Subsequently, it selects the next token in the output based on these probabilities. Temperature is a hyperparameter used to regulate the creativity level of LLMs in text generation. By adjusting the temperature value, the probability distribution of LLMs can be affected, resulting in more concentrated or diversified text output.

\subsection{Software developing}
Once an instance is generated, ChatDev uses it as a blueprint to guide the software development process. ChatDev systematically progresses through each phase detailed in the instance, orchestrating the collaboration of multiple agents to fulfill the tasks. Prior to executing a task, each agent is assigned a specific role, which may change depending on the phase. As illustrated in Figure \ref{fig2}, during the Designing phase, for example, ChatDev assigns two agents the roles of CEO (chief executive officer) and CTO (chief technology officer). These agents then work together through the Chat Chain to complete the design tasks. Throughout their dialogue, ChatDev incorporates three key mechanisms: Role Assignment, Memory Stream, and Self-Reflection, to ensure that the software development process is both efficient and accurate. ChatDev supports a total of seven roles: CEO (chief executive officer), CPO (chief product officer), CTO (chief technology officer), Programmer (also known as Coder), Reviewer, Designer, and Tester. Additionally, ChatDev offers a diverse set of software runtime environments, including numpy, matplotlib, pandas, tkinter, pillow, and flask, among others. Once the software development is complete, testers compile the software within the provided runtime environment and assess whether the instance used to guide the development was successful. These records serve as environmental feedback for the LLMs, facilitating continuous improvement and learning.

\subsection{Instance filtering}
Once the software development stage is completed, the instances are logged in an instance pool. As detailed in Table \ref{tab1}, each entry in the pool primarily consists of: the specific content of the instance, the frequency of occurrence of the instance, and the success count of instances that successfully passed compilation.

During the instance generation process guided by prompts, LLMs are susceptible to what we term ``instance hallucinations.'' Instance hallucinations refers to the occurrence of one or more of the following anomalies: 1) Incorrect phases. For example, the phase ``UsernameSet'' appears in the instance ``Designing $\to$ Coding $\to$ UsernameSet $\to$ Testing,'' but this phase is not defined in the phase explanations and does not have any specific meaning in software development; 2) Incorrect order. For example, the instance ``Designing $\to$ Testing $\to$ Coding'' indicates that testing is performed before coding, which is contrary to the standard sequence of software development. 3) A combination of both incorrect phases and incorrect order. Clearly, instances containing such irregularities have the potential to lead to failures in subsequent software compilation. To address this issue, we employ two mechanisms: the \textit{Replay mechanism} and the \textit{Heuristic filter}, to reduce the impact of instance hallucinations.

\begin{table}[htbp]
\caption{Examples for illustrating records within the instance pool. }
  \centering
  \begin{tabular}{cccc}
    \toprule
    ID & Instance & Frequency & Success Counts \\
    \midrule
    1 & C $\to$ T & 14 & 6 \\
    2 & D $\to$ C $\to$ T & 21 & 13 \\
    ··· & ··· & ··· & ···  \\
    n & D $\to$ R $\to$ C $\to$ T & 30 & 21 \\
    \bottomrule
  \end{tabular}

\label{tab1}
\end{table}

\textbf{Replay mechanism}: To leverage the experience gained from prior development efforts and enhance the efficiency of software development, we maintain a comprehensive record of every instance used to guide development in the instance pool. However, given the occurrence of hallucination phenomena, some successfully compiled instances may be instance hallucinations, and some failed instances may also be correct. Therefore, it is necessary to randomly select instances from the instance pool for re-evaluation. This process is also called ``replay''.
    \begin{equation}
        UCT(i)  = 1 - \frac{Q(i)}{N(i)} + C*\sqrt{\frac{\ln_{}{N} }{N(i)}} 
        \label{eq1}
    \end{equation} 
Where, $Q(i)$ represents the successful compilation count of the $i-th$ instance.
$N(i)$ represents the occurrence frequency of the $i-th$ instance.
$C$ is a constant used to balance success rates and occurrence frequency, defaulting to 1. $N$ represents the sum of occurrence frequency of all instances.

\textbf{Heuristic filter}: After replaying, the frequency and success count of each instance in the instance pool are updated. At this point, we can use heuristic rules to extract instances with high \textit{success rate (SR)} as the experience for LLMs:
    \begin{equation}
        SR(i) = \frac{Q(i)}{N(i)} 
        \label{eq2}
    \end{equation}
It should be noted that \textit{Heuristic filter} is to filter those instances with high success rate for experience gathering, while the focus of the \textit{Replay mechanism} is to select those instances with lower success rate and lower frequency of occurrence in the instance pool for testing, thereby reducing the impact of instance hallucinations on the entire software.

\subsection{LLM enhancing}
The final stage of our framework is LLM enhancing, where we mine process models from instances with a high success rate. The objective is to improve the dynamic instance generation capabilities of LLM by translating these process models into a textual description that is more accessible and comprehensible for LLM to utilize as prompt input.

Process mining has garnered widespread attention in the field of business process management in recent years \cite{van2012process}. This is due to its ability to accurately construct business process models from vast amounts of event logs and guide the step-by-step execution of business processes. In this paper, we opt to employ the Inductive Miner (IM) algorithm \cite{leemans2014discovering} to mine process model from instances, representing it in the form of BPMN (Business Process Model and Notation). The main reason for this choice is that IM excels in all performance metrics, as highlighted in the preceding Related Work section. BPMN is widely embraced in both industry and academia for its rich set of elements that enable the creation of simple and understandable process models, accurately expressing the user's intentions \cite{BPMN}. In the Table \ref{tab2}, we list the main elements covered in this paper, including start and end events, sequence flow, activities and gateways, where activities representing steps or phases in task execution, and gateways representing relationships between these steps.


\begin{table}[htbp]
\caption{The BPMN elements and descriptions. }
  \centering
  \begin{tabular}{c}
  \includegraphics[scale=0.32]{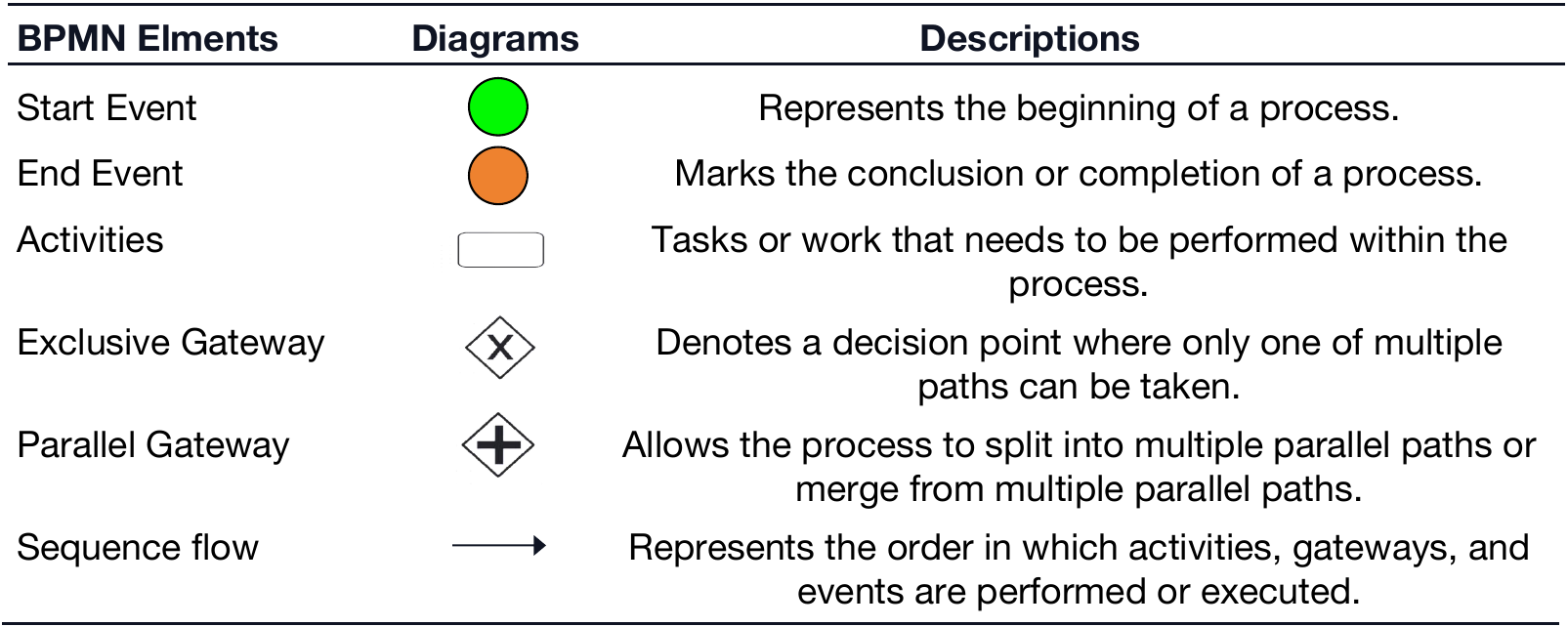} 
  \end{tabular}
\label{tab2}
\end{table}

\textbf{Definition 1} (directly follows relation $\twoheadrightarrow $). Let $ \sum $ be an alphabet such that $\top$ $\notin$ $\sum$ and $\bot$ $\notin$ $\sum$ and let $L$ be a language over $\sum$. We define the following relations:

\vspace{6pt}
\hspace{1.5cm} $a$ $\twoheadrightarrow$ $b$ $\Leftrightarrow$ $\exists_{\text{t}\in L}$ $t$ = $<\dots$ $ a,b,$ $\dots$$>$
\vspace{6pt}

\hspace{1.5cm} $\top$ $\twoheadrightarrow$ $a$ $\Leftrightarrow$ $\exists_{\text{t}\in L}$ $t$ = $<a,$ $\dots$$>$
\vspace{6pt}

\hspace{1.5cm} $a$ $\twoheadrightarrow$ $\bot$ $\Leftrightarrow$ $\exists_{\text{t}\in L}$ $t$ = $<\dots$$,a>$
\vspace{6pt}

\begin{figure}[htbp]
\centering
\centerline{\includegraphics[scale=0.77]{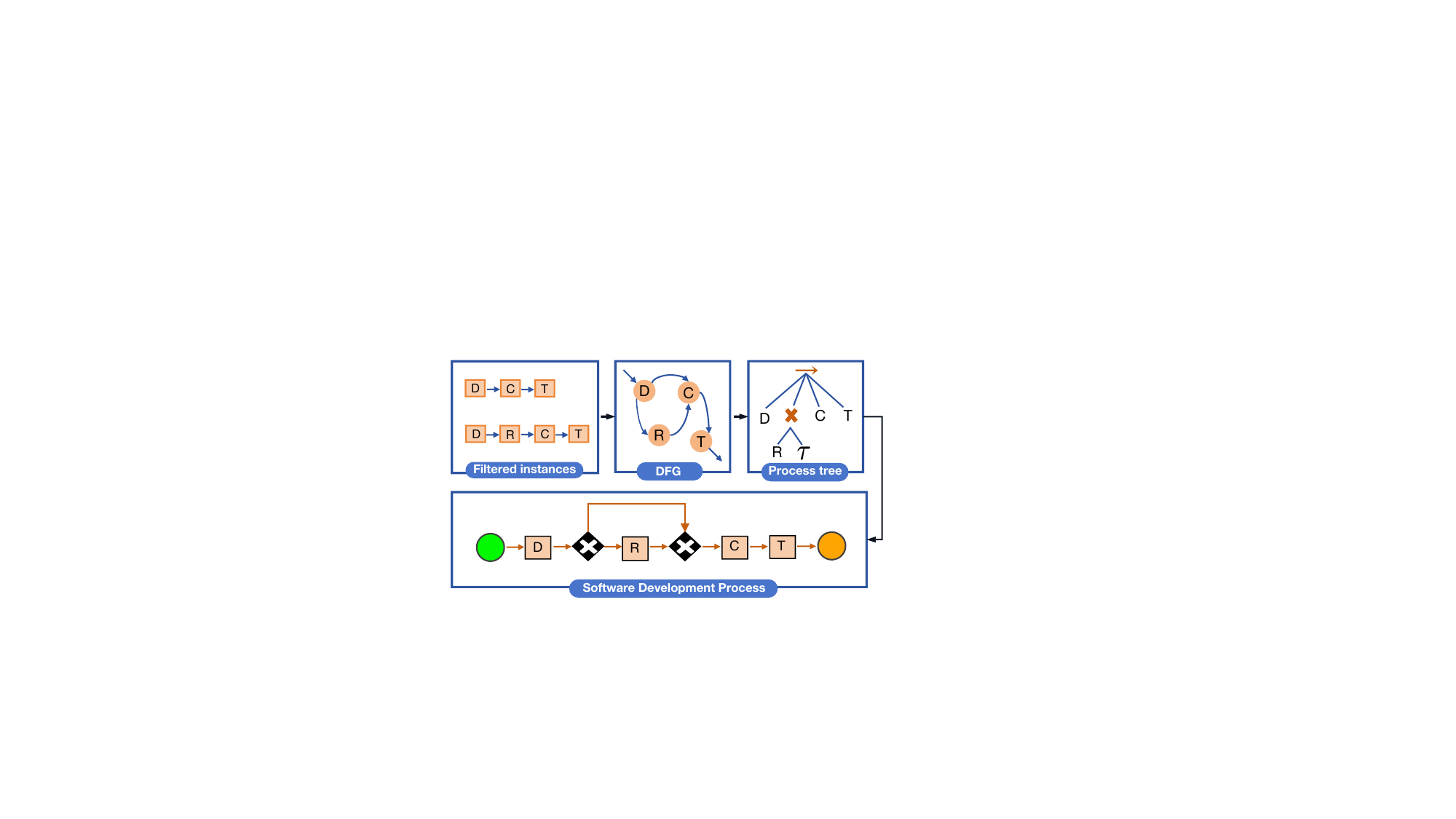}}
\caption{An example for illustrating process model mining.}
\label{fig5}
\end{figure}

Definition 1 states a relationship between two phases, where \textit{t } represents an instance, \textit{a} or \textit{b} represent phase, $\top$ and $\bot$ represent the start event and end event, respectively. The directly follows graph (DFG) is a directed graph: its nodes are phases, and its edges denote which phases can directly follow one another. In Figure \ref{fig5}, we provide an example illustrating how to mine the software development process model from two instances based on the IM algorithm. First, a directly follows graph is constructed from filter instances based on directly follows relation. Second, the IM algorithm transforms DFG into a process tree based on four \textit{cut operations}: sequence ($\to$), exclusive ($\times$), parallel ($+$), and loop ($\circlearrowleft$). The process tree is a directed connected graph without cycles. A node in the graph is either a \textit{branch node} or a \textit{leaf node}. Each leaf node represents phase from the collection of instances.

Each branch node, or operator node, has one or more children. These children can be other operator nodes or leaf nodes. It is worth noting that an invisible activity $\tau$ is often added in process tree to maintain the soundness. Finally, the process tree is equivalently transformed into BPMN to graphically display the software development process model from the instances. And the BPMN shows that the IM algorithm accurately mines selection relationships within dynamic instances by constructing the directly follows graph and conducting cut operations.

\begin{figure}[t]
\centering
\centerline{\includegraphics[scale=0.6]{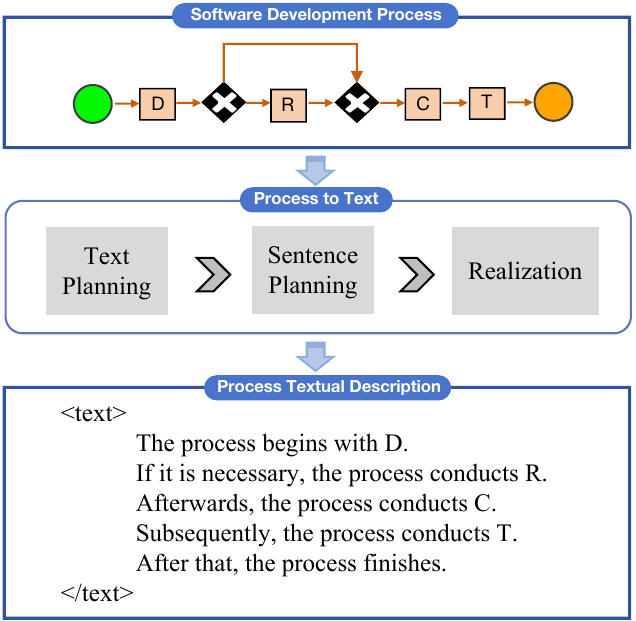}}
\caption{Conversion steps from software development process to process textual description.}
\label{fig6}
\end{figure}

Understanding BPMN can be challenging, especially since there are pairs of exclusive and parallel gateways in the model. Due to the emergence of different gateways, the relationship between activities in the model becomes more complicated, including sequence, concurrency and selection. Therefore, if a BPMN model is directly input into LLM, there’s no guarantee that LLM will accurately interpret the relationships between different activities within the model. Well, the impressive performance of LLM in various benchmark tests in recent years demonstrates their ability to comprehend the semantic information within text \cite{chang2024survey,zhao2023survey}. To bridge this gap, we first translate the BPMN model into a process textual description, as illustrated in Figure \ref{fig6}. This conversion involves three key steps: 1) \textit{Text Planning}: This initial step involves deciding what information needs to be communicated and establishing the sequence in which it should be presented. 2) \textit{Sentence Planning}: In this step, we select the appropriate vocabulary to convey the information that was identified in the previous step. 3) \textit{Realization}: The last step is to convert the acquired information into grammatically correct sentences.

\begin{figure}[htbp]
\centering
\centerline{\includegraphics[scale=0.44]{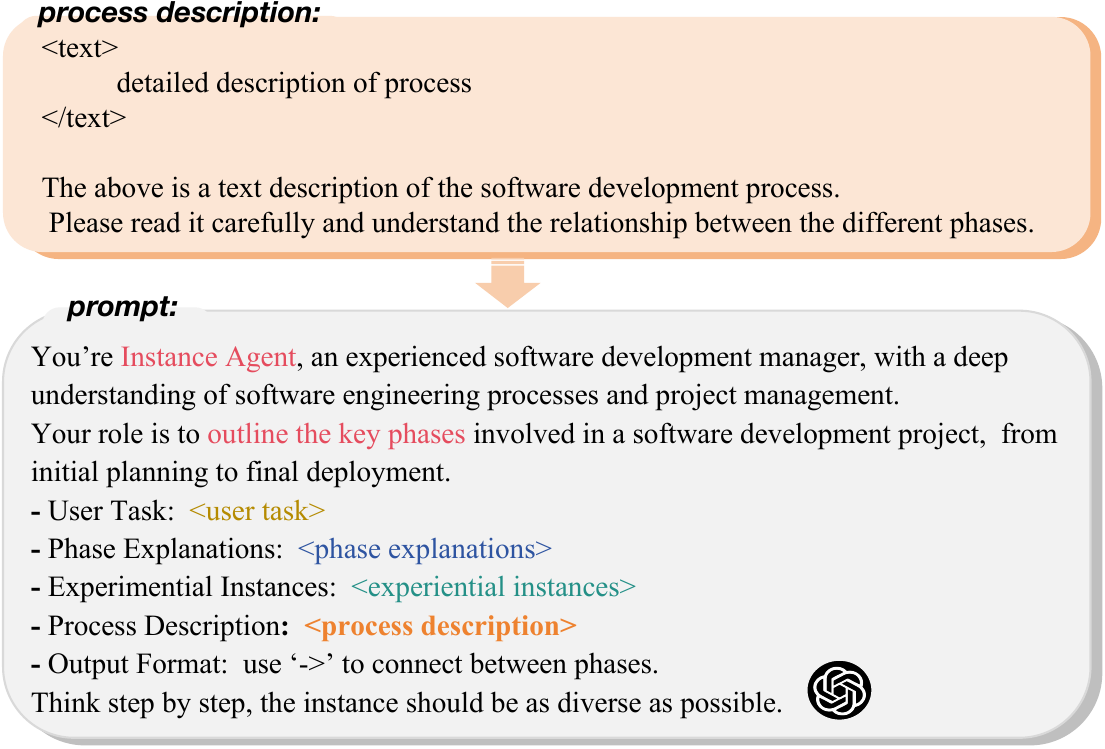}}
\caption{Prompt contents after adding process description.}
\label{fig14}
\end{figure}

Once the conversion is complete, we obtain a process textual description that retains the original semantics of the BPMN model. This description is then integrated into the prompt to enhance the dynamic instance generation capabilities of LLMs. The updated prompt, which includes the process textual description, is depicted in Figure \ref{fig14}.

\begin{table*}[b]
\caption{Software Development Lifecycle (SDLC) and Specific Phases}
  \centering
  \begin{tabular}{cc}
    \toprule
    SDLC & Specific Phases\\
    \midrule
    Software Requirements & DemandAnalysis,LanguageChoose \\
    Software Design &  DesignReview\\
    Software Construction & Coding, CodeComplete, Annotation, CodeConclusion, CodeReviewComment\\
    Software Quality &  CommentJudgement, CodeReviewModification, TestErrorSummary, TestModification\\
    Software Maintenance & EnvironmentDoc, Manual\\
    \bottomrule
  \end{tabular}
  
\label{tab2}
\end{table*}

\section{EXPERIMENTS}



\subsection{Setup}
Our experiment utilized OpenAI's GPT-3.5-turbo model\footnote{https://openai.com/} as the backbone for simulating multi-agent software development. For the Python-based system, we employed Python 3.9.18 as the external compiler interpreter, and all the code is publicly available\footnote{https://github.com/Aizhouym/Think-on-Process}.

\subsubsection{Datasets:}
To validate the feasibility of our framework,  we chose to use the NLDD dataset \cite{qian2023communicative}, an open source dataset dedicated to the field of ``Natural Language to Software Generation'', and the dataset contains more than 1200 software task prompts, these prompts cover the five major areas of Creation, Game, Education, Work, and Life. By selecting software tasks from these five categories with equal probability as user demand input, our experiment accounts for the diverse software development needs across various fields.

\subsubsection{Configurations:}
During the software development of ToP, we use ChatDev\footnote{https://github.com/OpenBMB/ChatDev} as the backbone. After generating an instance based on different user requirements, this instance is divided into multiple phases according to the software development lifecycle for execution. Each phase requires the participation of agents with different roles. \cite{shafiq2021literature} refined the software development lifecycle into five different categories based on the knowledge areas mentioned in SWEBOK \cite{abran2001guide}: 1) software requirements, 2) software design, 3) software construction, 4) software quality, and 5) software maintenance. To achieve more efficient software development, we further subdivided these five categories, resulting in a default instance with 14 phases, as shown in Table \ref{tab2}.

\subsection{Hyperparameter Analysis}
Large language models can typically control the randomness of their output by adjusting the temperature parameter. Generally, the temperature value falls within the range [0.0, 1.5]. A higher value indicates greater randomness and creativity, but it may also lead to more errors \cite{zhu2023hot}. Conversely, a lower value results in reduced randomness, generating content that is overly conservative and repetitive. 

As the temperature value decreases, the diversity instances decrease.  However, when the temperature value increases, a large number of irrelevant instances will be generated, resulting in a waste of learning resources.  Therefore, we need to explore a balance that allows LLMs to generate diverse instances while ensuring a reasonable success rate.  Before running the experiment, we would like to introduce the diversity metrics:

\begin{itemize}
    \item \textit{Diversity}: The diversity of each instance is quantified by comparing its phases and order variations against a default instance, which is composed of 14 predefined phases based on expert knowledge. A higher diversity value signifies a greater differences between the instances and a richer variety in the generated instances. Conversely, a lower diversity value suggests a lesser degree of variation, with instances tending to be more similar or invariant \cite{menezes2018diversity}.

        \begin{equation}
            Diversity(i) = \frac{Change(Phases) + Change(Orders)}{2}
            \label{eq3}
        \end{equation}

        \begin{equation}
            Change(Phases) =\frac{\lvert P - (P \cap N)\rvert  + \lvert N - (P\cap N)\rvert}{\lvert  P \cup N \rvert }
            \label{eq4}
        \end{equation}
        
        \begin{equation}
            Change(Orders) =\frac{\lvert O - (O \cap E)\rvert  + \lvert E - (O\cap E)\rvert}{\lvert  O \cup E \rvert } 
            \label{eq5}
        \end{equation}

     In equation \ref{eq3}, diversity of instance $i$ is calculated by computing the average of the changes in phases and orders compared to the default instance. In equation \ref{eq4}, $P$ represents the set of phases constituting the default instance, and $N$ represents the set of phases constituting instance $i$. In equation \ref{eq5}, $O$ represents the set of orders constituting the default instance, and $E$ represents the set of orders constituting instance $i$.
    
\end{itemize}

\begin{figure}[htbp]
\centering
\centerline{\includegraphics[scale=0.23]{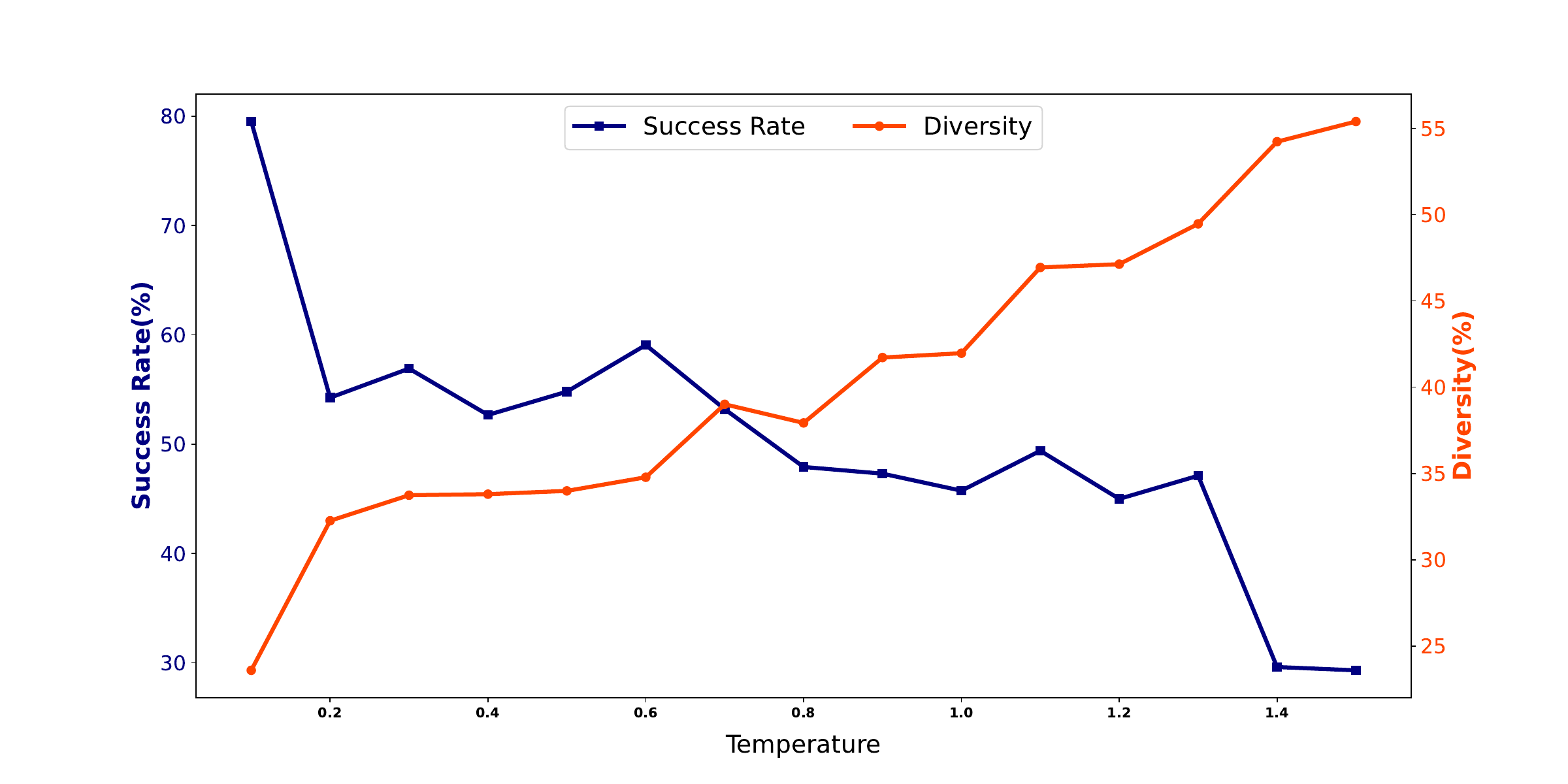}}
\caption{The impact of different temperature on \textit{SR} and \textit{Diversity}.}
\label{fig7}
\end{figure}

Figure \ref{fig7} illustrates the relationship between temperature and two metrics: success rate and diversity. When the temperature increases from 0.0 to 0.2, the success rate experiences a substantial decline of 25.7\%. Concurrently, diversity shows a marked increase, rising by 9.6\%. As the temperature fluctuates between 0.2 and 0.6, there is a minimal change in both success rate and diversity, suggesting a relatively stable condition. Beyond a temperature of 0.6, however, there is a notable decrease in success rate accompanied by a significant rise in diversity. For instance, at a temperature of 1.4, over half of the instance contents are randomly generated, and the success rate falls below 30\%, indicating that these instances are almost impossible to successfully guide software development.

In summary, as the temperature rises, the diversity of instances produced by LLMs increases. Conversely, the success rate of these instances in subsequent software development diminishes. Therefore, to ensure that instances generated by LLMs exhibit increased diversity while still achieving a reasonable level of success in test compilation, the temperature should not be set too high (resulting in low success rate) or too low (leading to low diversity). An intermediate temperature value is often the most effective choice. In this study, we default to a temperature of 0.6.

\subsection{UCT Analysis}
The occurrence of instance hallucinations can lead to instances that contain errors when compiled within the external environment (for example, the wrong instance may pass the compilation, but the correct one cannot pass). This, in turn, may result in the incorrect mining of the software development process model. To address this issue, we implement a replay mechanism that selects instances from the instance pool for re-evaluation. The choice of strategy significantly impacts the selection results. Common strategies include:

1) \textit{Failure rate-based}: This strategy gives priority to those instances with higher failure rates, hoping to reduce the number of instances that are semantically correct but fail to compile. 2) \textit{Frequency-based}: This strategy preferentially selects instances with low access times, because those instances with low frequency cannot determine whether they are instance hallucinations. 3) \textit{UCT (Upper Confidence Bound for Trees)}: This strategy is an overall consideration of instance failure rate and access frequency, seeking to achieve a balance between these two factors.

\begin{figure}[t]
\centering
\centerline{\includegraphics[scale=0.34]{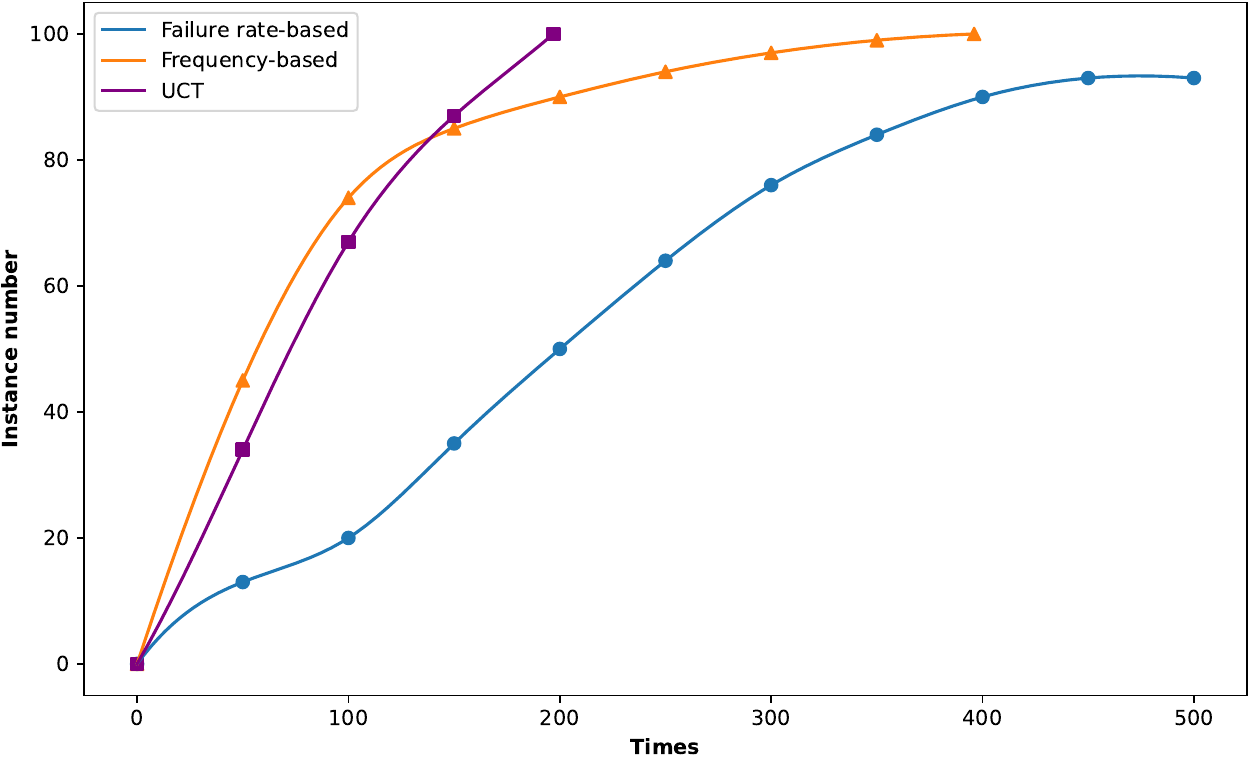}}
\caption{Comparison of different selection strategies.}
\label{fig8}
\end{figure}

To assess which instance selection strategy is most effective for the replay mechanism, we manually selected 100 diverse instances (including hallucination instances) from the instance pool which containing 1000 instances. We then employed three different selection strategies for replay and evaluated required attempt times for each strategy to complete all instance traversals. It’s important to note that each instance retained its frequency and success rate attributes when it was initially selected from the instance pool. As depicted in Figure \ref{fig8}, the Frequency-based strategy initially quickly accesses instances with smaller times. However, as the number of times increases, the available instances for selection decrease, making it challenging to access instances with high initial frequency. In contrast, the UCT strategy may not perform as well at the outset but demonstrates strong access stability, ensuring continuous access to new instances. After 197 times, the UCT strategy successfully traversed all instances. The least efficient strategy is Failure rate-based, because even if the Times = 500, it is still impossible to complete the traversal of all instances. This is because if there are instances with success rate = 100\% among 100 instances, these instances will never be accessed.

To sum up, UCT as an instance selection strategy for the replay mechanism is relatively reasonable and reliable. It effectively combines elements of both the Failure rate-based and Frequency-based strategies, providing a balanced approach to refining the instance pool.

\subsection{Filtering Mechanism Effectiveness}
\begin{figure}[h]
\centering
\centerline{\includegraphics[scale=0.34]{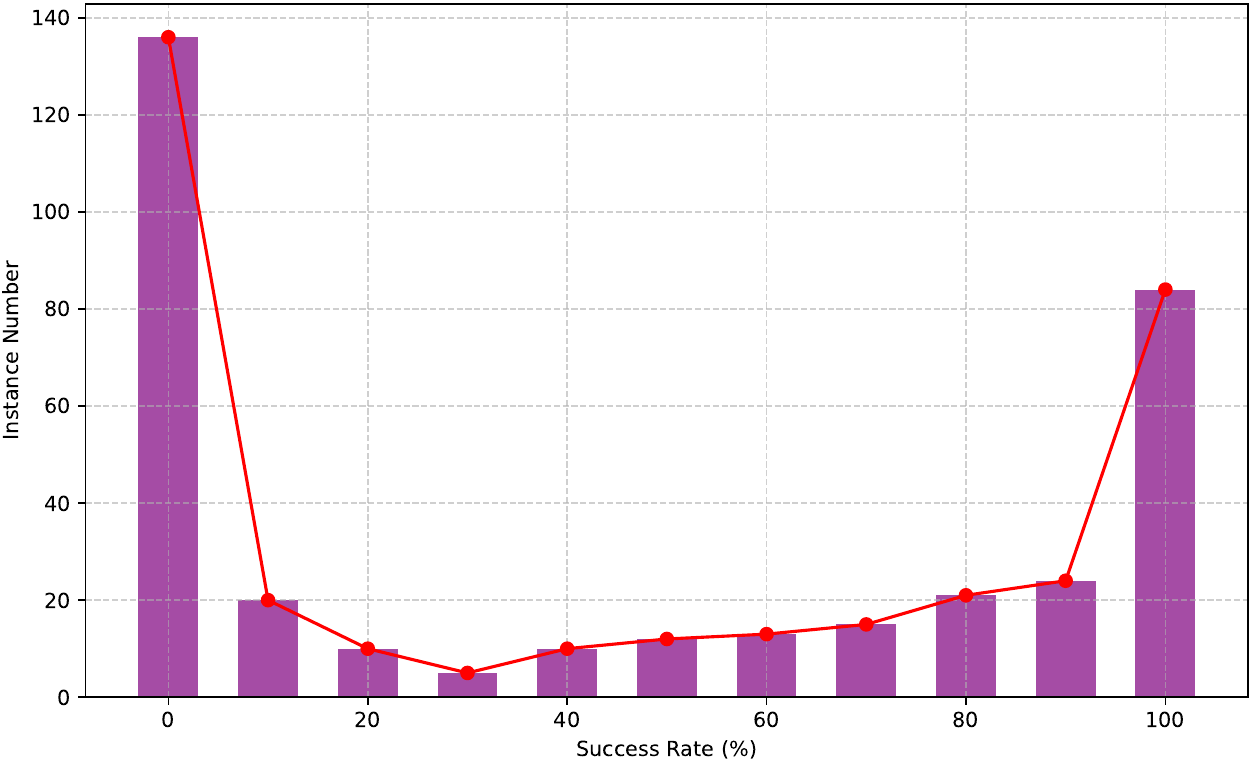}}
\caption{The number of process instances under different success rates.}
\label{fig9}
\end{figure}

From LLMs generating instances, to multi-agent collaboration for software development, to subsequent completion of external environment compilation, the entire process takes about 8 minutes. In this article, we generated 1,000 instances and stored 350 different variants (i.e., diverse instances) in the instance pool. We counted the number of instances obtained under different success rate filters, and the distribution of results is shown in Figure \ref{fig9}. The data reveals that the majority of instances generated by LLMs are not compatible with the external environment. This outcome is understandable, given that software development under LLMs guidance involves many independent adjustments, which undoubtedly increase the likelihood of errors. Humans, on the other hand, can adjust the development process in real-time based on past experience and external feedback.

At the same time, it can be observed that the line chart presents a bimodal distribution, with the inflection point located at 30\%. Therefore, we choose a success rate of 30\% as the threshold for instance filtering to reduce the impact of instance hallucinations.

\begin{figure*}[h]
\centering
\subfigure[The process model without filtering.] { \label{fig10:}
\includegraphics[scale = 0.08]{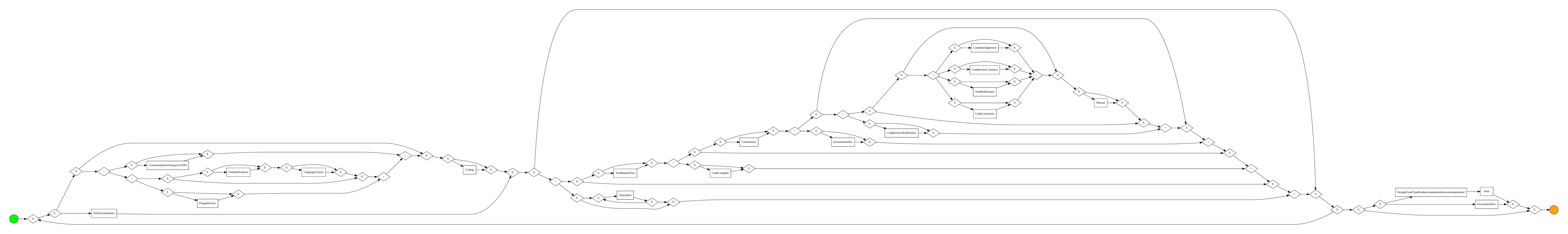}
}
\subfigure[The process model based on filtered instances.] { \label{fig11:}
\includegraphics[scale = 0.126]{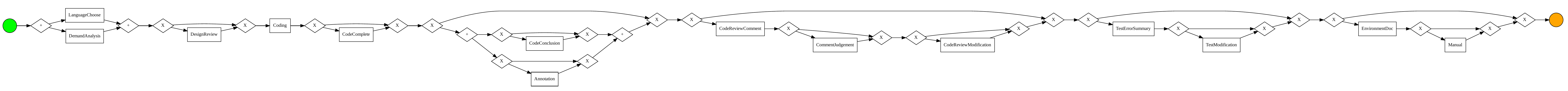}
}

\caption{The process models of software development for multi-agent system.}
\label{fig}
\end{figure*}

\begin{figure*}[b]
\centering
\subfigure[GPT-3.5-turbo] { \label{fig12:}
\includegraphics[scale = 0.48]{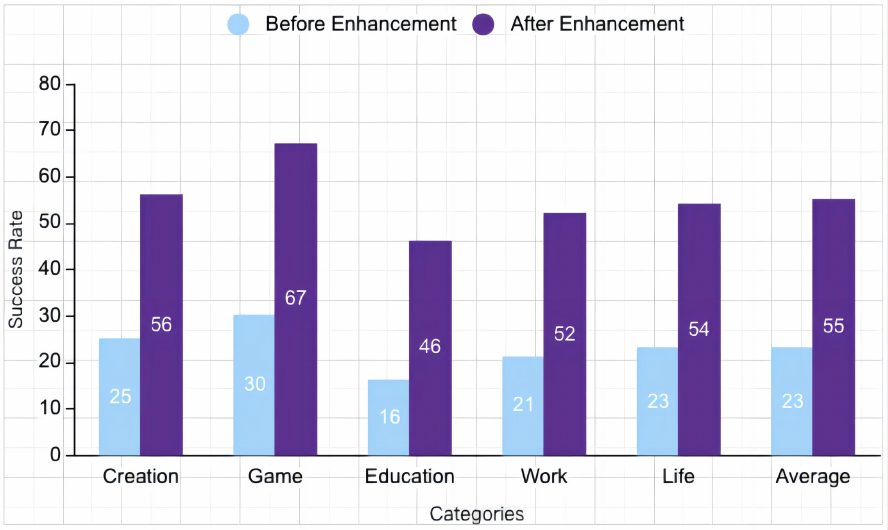}
}
\subfigure[GPT-4] { \label{fig13:}
\includegraphics[scale = 0.48]{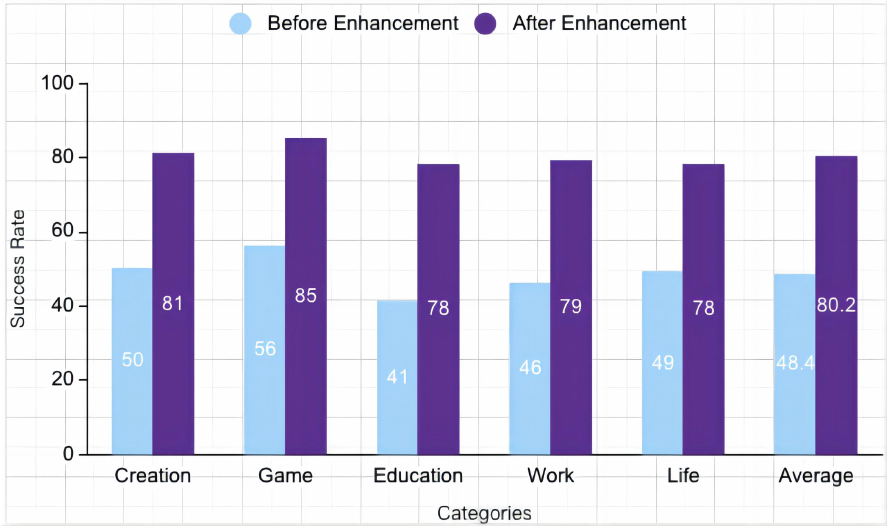}
}

\caption{Effectiveness of ToP's enhancements to LLMs on 5 different types of development tasks.}
\label{figx}
\end{figure*}

In order to clearly compare the differences in software development process models before and after filtering, we use the IM algorithm to mine the models of the instance pools of the two situations respectively. The results are presented in Figure \ref{fig}.

Figure\ref{fig10:}, the software development process obtained directly without filtering is chaotic. The model includes 20 phases (here we regard each activity as a phase in software development) and 14 parallel gateways and 58 exclusive gateways, and there are obvious errors in the phases (e.g., ``UsernameSet'', ``ExternalReview'', ``Comments'', ``EmailSet'', ``EmploymentDoc'') and orders (e.g., ``DesignReview $\to$ DemandAnalysis'', ``Coding $\to$ CommentJudgement'', ``LanguageChoose $\to$ Manual'', ``Annotation $\to$ TestErrorModification'', ``CodeReviewComment $\to$ DesignReview'') in the process model. In contrast, Figure \ref{fig11:} presents a clear software development process. The number of phases is reduced to 14, aligning with the phases outlined in our experimental phase explanations. Additionally, the number of gateways is reduced by 65\%, with only 4 parallel gateways and 24 exclusive gateways, indicating a more efficient and coherent process model.

\subsection{Enhancement Experiment}
The software development process model can serve as the accumulated development experience for LLM, enhancing their dynamic process generation capabilities. By converting the process model into a process textual description with equivalent semantics, we can input this description to LLM as a prompt. To assess the effectiveness of this approach, 200 different software development tasks were selected from each of the five major categories of the NLDD dataset as user input. The success rates of the software development process under the two methods (with and without process textual description input) were then compared.

From Figure \ref{fig12:}, it is evident that the incorporation of process textual description directly impacts the software development success rate of LLM. The development success rate of LLM is elevated by an average of 32\% across the five distinct software development categories when compared to the baseline without enhancement. Notably, the category with the highest success rate is Game, with a leap from 30\% to 67\% after LLM enhancement. Conversely, the category with the lowest success rate is Education, where the enhancement brings the success rate up from 18\% to 46\%. This phenomenon is logical, as LLM possess a deeper understanding of simple codes and can readily write and call functions. Education-related software development, however, relies heavily on specific professional domain knowledge, posing a challenge for general language models. Figure \ref{fig13:} illustrates the results obtained with GPT-4. Due to the promotion of the model's own performance, the success rate across all categories has increased no matter existing the enhancement or not. In coparision to different categories, it is noteworthy that the highest success rate before enhancement is only 56\%. Post-enhancement, the lowest success rate across all categories reaches 78\%, while the highest reaches 85\%, highlighting a substantial improvement.

In conclusion, integrating process textual descriptions with equivalent semantics into LLM can substantially bolster their generation capabilities. This approach enables the acquisition of more robust and comprehensive development software.

\section{RELATED WORK}
\textbf{Software Development }Software development is crucial for human utilization of automated data processing. It should be carried out gradually and in phases, involving continuous user participation in the design and programming processes \cite{poppendieck2012lean,despa2014comparative,stober2010best}. An efficient and rational software development process can guide developers in swiftly accomplishing intricate development tasks. Bassil et al. \cite{bassil2012simulation} proposes a novel waterfall model that effectively mitigates potential issues in the software development process, such as budget overruns, delivery delays, and customer dissatisfaction. Additionally, in various stages of software development, people are increasingly adopting Al technologies to assist and expedite the software development process \cite{wan2022automated, fan2023large,white2023chatgpt}. For instance, during the requirements phase, Pudlitz et al. \cite{pudlitz2019extraction} suggests the use of a self-trained named-entity recognition model employing bidirectional LSTM and CNN to extract the status of natural language requirements. And during the test phase, unlike LLMs, which is used for the entire unit test \cite{lukasczyk2022pynguin}, Lemieux et al. \cite{lemieux2023codamosa} combined the traditional search-based software testing technology with LLMs technology. When the test coverage of the traditional technology reaches the maximum, LLMs generates corresponding test cases for the functions that are not covered. However, the software processes employed in the aforementioned studies are static, neglecting the consideration of dynamic software development processes.

\textbf {Large Language Models} Large Language models such as ChatGPT \cite{ChatGPT} and LLaMA \cite{touvron2023llama} have demonstrated their remarkable capabilities in the field of natural language processing, after extensive training on massive datasets \cite{kaplan2020scaling,wang2023voyager,chen2023agentverse,schick2024toolformer}. Qin et al. \cite{qin2023large} proposed a new sorting prompt technique (Pairwise Ranking Prompt) that has achieved significant improvements on existing datasets. Li et al. \cite{li2023making} leveraged the outstanding ability of large language models in semantic understanding to propose LLaRA, which cleverly transforms large models into dense retrieval encoders, greatly improving the performance of the model on various dense retrieval benchmarks. Song et al. \cite{song2023llm} proposes using LLMs for few-shot planning and enhancing LLMs with physical principles to generate and update grounded plans in the current environment. Sharan et al. \cite{sharan2023llmassist} introduces a novel hybrid planner that combines traditional rules with LLMs. It utilizes the reasoning capabilities of LLMs for task planning, achieving state-of-the-art performance on the nuPlan benchmark. Planing technology enables LLMs to efficiently choose different processes to accomplish tasks, providing a foundational guarantee for the generation of dynamic processes.

Recently, there has been a growing interest in developing software with LLMs \cite{roziere2023code,luo2023wizardcoder,macneil2023experiences}. CodeGen \cite{nijkamp2022codegen} can generate code for user-defined tasks step by step through multi-turn dialogues with LLMs. ChatDev \cite{qian2023communicative} divides software development into four steps and guides the collaboration of multi-agent through these steps to accomplish the development of applications and the generation of documentation. However, LLMs in the process of software development are prone to experiencing hallucination phenomena. Lightman et al. and Mackenna et al. \cite{Lightman2023LetsVS,McKenna2023SourcesOH} believed that the generation of hallucinations in LLMs is caused by overfitting or noisy data during the training process. In summary, LLMs can effectively generate dynamic processes, but they are prone to hallucination phenomena, which may hinder their effectiveness in guiding software development. 

\textbf{Process Mining} Process mining is a data-driven technique that can be used to discovery, analyze, and improve the ongoing process model based on historical records (e.g., massive instances) \cite{van2012process}. Process discovery is one of the three major scenarios in process mining, which is to automatically mine process models from historical records without any apriori information \cite{wilprocessmining}. Until now, many process discovery algorithms have been proposed like Alpha \cite{van2004workflow}, Genetic Miner \cite{van2005genetic}, Inductive Miner \cite{leemans2014discovering}, etc. It is crucial to emphasize that process discovery algorithms not only showcase historical behaviors but also reveal additional behaviors beyond the historical scope. In other words, the process model's behavior may surpass historical behavior, constituting what is commonly referred to as the generalization \cite{van2016unified} ability in the field of process mining. Essentially, the extra behaviors derived from historical records can be leveraged to enhance the capabilities of LLMs. Furthermore, fitness \cite{berti2019reviving} and precision \cite{adriansyah2013alignment} serve as assurances to prevent the process model from deviating from reality. In this setting, Inductive Miner (IM) is selected as the mining algorithm due to its commendable performance in generalization, fitness, and precision \cite{augusto2018automated}. Moreover, the resulting process models from IM are structured, facilitating readability and comprehension. Though BPMN models represent complex relationships between different activities by exclusive gateways, parallel gateways, swimlanes, etc. It is difficult for most people to understand BPMN models because it often requires corresponding expert knowledge. The emergence of BPM2Text technology solves this problem well \cite{leopold2012generating,rodrigues2014tool,rodrigues2016bpm2text,delicado2017nlp4bpm}, by converting BPMN models into natural language texts with the same semantics, so that non-professionals can also understand and verify these models well.

\section{CONCLUSIONS}
The automation of software development is a valuable research topic that can effectively enhance team efficiency. In this paper, we propose a framework called ToP for dynamically generating software development processes, which can produce various instances to guide multi-agent in software development tasks. To start with, LLMs would generate an instance based on user requirement, then the instance be used to guide multi-agent for software developing. Furthermore, we utilize compilers as external tools for feedback on these instances, employing heuristic algorithm to mitigate the phenomenon of hallucination in the generated instances. Finally, we apply process mining techniques to extract process models from successful instances as experiential knowledge. ToP leverages experiential knowledge to enhance the ability of LLMs to generate other instances.  Extensive experiments have been conducted on two LLMs, GPT-3.5 and GPT-4, demonstrating that our approach effectively improves the success rates of software development across five categories of tasks.

In the future, we will continue to explore the following aspects: on one hand, identifying more comprehensive feedback mechanisms to replace compilers; on the other hand, fully leveraging failed instances to further enhance LLMs, rather than limiting our focus to successful instances only.

\bibliography{cited}
\bibliographystyle{ACM-Reference-Format}
\end{document}